\documentclass[11pt,a4paper]{article}
\usepackage{jheppub} 
\usepackage{mathrsfs}
\usepackage{amsmath}
\usepackage{amssymb}
\usepackage[mathscr]{euscript}

\title{A note on chiral trace relations from qq-characters}

\author[a,b]{Saebyeok Jeong}
\author[b]{and Xinyu Zhang}

\affiliation[a]{C.N. Yang Institute for Theoretical Physics, Stony Brook University,\\ Stony Brook, NY 11794-3840, USA}
\affiliation[b]{NHETC and Department of Physics and Astronomy, Rutgers University, \\ Piscataway, NJ 08854, USA}

\emailAdd{saebyeok.jeong@physics.rutgers.edu}
\emailAdd{zhangxinyuphysics@gmail.com}

\abstract{
We study chiral operators in four-dimensional $\mathcal{N}=2$ supersymmetric
gauge theories. We provide a general procedure to derive the exact
relations among the vacuum expectation values of chiral operators
in the $\Omega$-background using the non-perturbative Dyson-Schwinger
equations of qq-characters. We demonstrate our procedure using several
examples. For pure $\mathrm{SU}(N)$ gauge theory and $\mathrm{SU}(N)$
gauge theory with $2N$ fundamental hypermultiplets, we verify the
relations conjectured previously in the literature. We also briefly
discuss the relations in $\mathrm{SU}(N)$ linear superconformal quiver
gauge theories.
}

\arxivnumber{1910.10864}

\begin{document}

\maketitle


\section{Introduction}

Ever since the pioneering work of Seiberg and Witten \cite{Seiberg:1994rs,Seiberg:1994aj},
four-dimensional $\mathcal{N}=2$ supersymmetric gauge theories have
been repeatedly investigated from various viewpoints. Both the low
energy effective prepotential $\mathcal{F}_{\mathrm{SW}}$ and the
correlation functions of $\mathcal{N}=2$ chiral operators can be
determined in terms of period integrals of the Seiberg-Witten meromorphic
differential $\lambda$ on the Seiberg-Witten curve $\Sigma$. 

In order to provide a rigorous derivation of the Seiberg-Witten solution,
it is useful to put the theory in the $\Omega$-background \cite{Moore:1997dj,Nekrasov:2002qd},
which effectively regularizes the infinite volume of the spacetime
$\mathbb{R}^{4}$ while preserving a part of the deformed supersymmetry.
The partition function $\mathcal{Z}$ in the $\Omega$-background
can be computed exactly using the localization technique for a large
class of traditional Lagrangian theories constructed using the vector
multiplet with gauge group $G$ and the hypermultiplet in the representation
$\mathfrak{R}$ of $G$. When the gauge group is a unitary group, the result can be written as a statistical
sum over a collection of Young diagrams $\vec{\lambda}$ labeling
the special instanton configurations \cite{Nekrasov:2002qd},
\begin{equation}
\mathcal{Z}\left(\vec{a},\vec{m},\mathtt{q};\varepsilon_{1},\varepsilon_{2}\right)=\sum_{\vec{\lambda}}\mu_{\vec{\lambda}},
\end{equation}
where $\vec{a}$, $\vec{m}$, $\mathtt{q}$, $\left(\varepsilon_{1},\varepsilon_{2}\right)$
are the Coulomb branch parameters, the masses of the hypermultiplets,
the instanton counting parameter, and the $\Omega$-deformation parameters,
respectively. The measure factor $\mu_{\vec{\lambda}}$ contains the
classical, the one-loop and the instanton contributions from both
the vector multiplet and the hypermultiplet. In the refined genus
expansion of the partition function around the flat space limit $\varepsilon_{1},\varepsilon_{2}\to0$,
\begin{equation}
\log\mathcal{Z}=-\sum_{g,n=0}^{\infty}\left(\varepsilon_{1}\varepsilon_{2}\right)^{g-1}\left(\varepsilon_{1}+\varepsilon_{2}\right)^{n}\mathcal{F}_{\left(g,\frac{n}{2}\right)},\label{eq:logZ}
\end{equation}
the leading term $\mathcal{F}_{\left(0,0\right)}$ coincides with
the low energy effective prepotential $\mathcal{F}_{\mathrm{SW}}$ of the theory
on $\mathbb{R}^{4}$. Hence we can derive the Seiberg-Witten geometry
directly for a large class of $\mathcal{N}=2$ theories \cite{Nekrasov:2003rj,Shadchin:2004yx,Shadchin:2005cc,Nekrasov:2012xe,Zhang:2019msw}.
The higher order terms in the expansion (\ref{eq:logZ}) compute the
couplings of the theory to the background gravitational field \cite{Nekrasov:2002qd,Nekrasov:2003rj},
and are naturally linked with topological strings on non-compact Calabi-Yau
threefolds. 

Another interesting relation between the partition function in the
$\Omega$-background and the disconnected partition function of A-model
topological strings on Riemann surfaces was discovered
in \cite{Losev:2003py} and later generalized in \cite{Marshakov:2006ii,Nekrasov:2009zza,Zhang:2016xqi}.
The higher Casimir operators in the gauge theory are mapped to gravitational
descendants of the Kahler form in the topological string theory. 

It was conjectured in \cite{Nekrasov:2002qd,Losev:2003py} that the
partition function in the $\Omega$-background should be related to
certain two-dimensional conformal field theories or deformations thereof.
A concrete realization of this conjecture was found between the four-dimensional
$\mathcal{N}=2$ theory constructed by compactifying the six-dimensional
$(2,0)$ theory on a punctured Riemann surface $\mathcal{C}$ \cite{Gaiotto:2009we,Gaiotto:2009hg}
and two-dimensional Liouville/Toda theory living on $\mathcal{C}$
\cite{Alday:2009aq,Wyllard:2009hg}. Meanwhile, if we take the limit
$\varepsilon_{1}\to0$ and keep $\varepsilon_{2}=\hbar$ fixed so
that the $\Omega$-background preserves two-dimensional $\mathcal{N}=2$
super-Poincare invariance, we get a quantization of the classical
algebraic integrable systems underlying the four-dimensional $\mathcal{N}=2$
theories \cite{Nekrasov:2009rc,Nekrasov:2011bc,Nekrasov:2013xda}. 

Therefore, we have sufficient reasons to carefully explore all the
consequences of the generic $\Omega$-background. In this paper, we
focus on the $\mathcal{N}=2$ chiral ring, which is believed to be
freely generated on $\mathbb{R}^{4}$. Hence, we can choose a finite-dimensional
basis of chiral operators such that any element of the chiral ring
can be represented uniquely as a polynomial in the basis elements.
The number of generators of the chiral ring is the complex dimension
of the Coulomb branch, and the polynomial equations representing generic
chiral operators in terms of the chosen basis elements are called
chiral ring relations. For simple Lagrangian theories on $\mathbb{R}^{4}$,
the chiral ring relations were derived in \cite{Cachazo:2002ry,Cachazo:2003yc}.
It is interesting to ask what happens when we introduce the $\Omega$-deformation.
The vacuum expectation value of a chiral operator $\mathcal{O}$ in
the $\Omega$-background is given by \cite{Losev:2003py,Marshakov:2006ii}
\begin{equation}
\left\langle \mathcal{O}\right\rangle =\frac{1}{\mathcal{Z}}\sum_{\vec{\lambda}}\mu_{\vec{\lambda}}\mathcal{O}_{\vec{\lambda}},\label{eq:vev}
\end{equation}
where $\mathcal{O}_{\vec{\lambda}}$ is the value of $\mathcal{O}$
evaluated at $\vec{\lambda}$, and the vacuum expectation value of
$\mathcal{O}$ on $\mathbb{R}^{4}$ can be obtained by taking the flat space limit,
\begin{equation}
\left\langle \mathcal{O}\right\rangle _{\mathbb{R}^{4}}=\lim_{\varepsilon_{1},\varepsilon_{2}\to0}\left\langle \mathcal{O}\right\rangle.
\end{equation}
For simplicity, we adopt the terminology used in \cite{Beccaria:2017rfz,Fachechi:2018bhe},
and refer to the equations representing the vacuum expectation values
of generic chiral operators in terms of the vacuum expectation values
of the chosen basis elements as \emph{chiral trace relations}. On $\mathbb{R}^{4}$,
the chiral trace relations contain the same amount of information
as the chiral ring relations due to the property of the cluster decomposition
(see (\ref{eq:cluster})). However, they are not equivalent once we
introduce the $\Omega$-deformation, since the $\Omega$-background
breaks the translational symmetry of $\mathbb{R}^{4}$. A priori it
is not guaranteed that chiral trace relations exist in the general
$\Omega$-background. Even if they exist, they need not to be polynomial
equations. Regardless, the chiral operators are conjectured to be
mapped to the integrals of motion in two-dimensional conformal field
theories \cite{Alday:2009aq,Alba:2010qc,Fateev:2011hq}. From the
considerations in the conformal field theory, we expect such chiral
trace relations in the $\Omega$-background still exist. Indeed, several
groups recently computed the vacuum expectation values of chiral operators
in the $\Omega$-background for some simple theories, and conjectured
the chiral trace relations in the $\Omega$-background based on the
explicit series expansion in $\mathtt{q}$ \cite{Fucito:2015ofa,Ashok:2016ewb,Beccaria:2017rfz,Fachechi:2018bhe}.
The problem is that neither the brute force method nor the analysis
based on two-dimensional conformal field theory can be easily generalized
to gauge theories with gauge group of high rank or with general matter content.

The purpose of this note is to provide a more systematic field theoretical
derivation of the chiral trace relations in the $\Omega$-background.
We will show that the non-perturbative Dyson-Schwinger equations,
which follow from the regularity of the vacuum expectation values
of qq-characters \cite{Nekrasov:2015wsu}, give a simple framework
for investigating the chiral trace relations. Our approach can be
regarded as the natural counterpart of the derivation of the classical
chiral ring relations on $\mathbb{R}^{4}$. We will work out the details
in the pure $\mathrm{SU}(N)$ gauge theory and the $\mathrm{SU}(N)$
gauge theory with $2N$ fundamental hypermultiplets to illustrate
our strategy, and then generalize to the linear $\mathrm{SU}(N)$
superconformal quiver gauge theories. We find that the chiral trace
relations in the $\Omega$-background can dramatically modify the
relations on $\mathbb{R}^{4}$. Similar strategy was also used in
the derivation of the BPZ equations from the point of view of four-dimensional
$\mathcal{N}=2$ supersymmetric gauge theories \cite{Nekrasov:2017gzb,Jeong:2017mfh}. 

It is possible to generalize our approach to other $\mathcal{N}=2$
theories if their qq-characters are known. For example, it is a good
exercise to derive the chiral trace relations in the $\Omega$-background
for the $\mathcal{N}=2^{*}$ theory, and check the conjectures made
in \cite{Ashok:2016ewb,Beccaria:2017rfz,Fachechi:2018bhe}. It is
also interesting to introduce half-BPS surface defects. There are
several different approaches to introduce surface operators in the
$\Omega$-background \cite{Nekrasov:2017rqy}, and it is straightforward to
derive the chiral trace relations for all
these approaches. Some of the cases have already
been studied in \cite{Jeong:2018qpc} in the context of their relation
to the symplectic geometry of the moduli space of flat connections
on a Riemann surface. Also, it was discovered in \cite{Jeong:2017pai}
that the surface defect partition functions split into parts at special
loci of the moduli space, accounting for the splitting of degenerate
levels in the corresponding quantum integrable system. It would be
interesting to study how the chiral trace relations are affected at
those loci. In \cite{Bourgine:2017jsi}, the gauge theory partition
functions were reconstructed in terms of representations of the quantum
toroidal algebra of $\mathfrak{gl}(1)$. The reconstruction was further
extended to the surface defect partition functions in \cite{Bourgine:2019phm},
with a proper generalization of the relevant quantum algebra. It would
be nice to investigate the algebraic meaning of the chiral trace relations
discussed in this note in this algebraic engineering context. Another
interesting problem is to consider the qq-characters when we put the
theory on compact spaces such as (squashed) four-spheres \cite{Pestun:2007rz,Hama:2012bg}. 

The rest of the paper is organized as follows. In section \ref{sec:strategy},
we outline our general strategy to derive the chiral trace relations
in the $\Omega$-background. In section \ref{sec:SU(N)}, we show
explicitly how to obtain the chiral trace relations in the $\Omega$-background
in the pure $\mathrm{SU}(N)$ gauge theory and the $\mathrm{SU}(N)$
gauge theory with $2N$ fundamental hypermultiplets. In section \ref{sec:quiver},
we generalize our discussion to the linear $\mathrm{SU}(N)$ superconformal
quiver gauge theories. 

\section{General strategy \label{sec:strategy}}

\subsection{Chiral ring relations on $\mathbb{R}^{4}$}

In four-dimensional $\mathcal{N}=2$ supersymmetric gauge theories,
a chiral operator $\mathcal{O}(x)$ is defined to be a gauge-invariant
local operator annihilated by the action of all supercharges $\bar{Q}_{\dot{\alpha}A}$
of one chirality,
\begin{equation}
\left[\bar{Q}_{\dot{\alpha}A},\mathcal{O}\left(x\right)\right]=0.\label{eq:chiral}
\end{equation}
Given such an operator, one can always generate its superpartners
by acting on it with the $Q_{\alpha A}$. It follows immediately from
the $\mathcal{N}=2$ supersymmetry algebra on $\mathbb{R}^{4}$ and
the Jacobi identity that a product of chiral operators is still chiral,
and its vacuum expectation value is independent of their spacetime
positions. Then one can take the limit of large separation and apply
the cluster decomposition to factorize correlation functions of chiral
operators,
\begin{equation}
\left\langle \mathcal{O}_{1}\left(x_{1}\right)\mathcal{O}_{2}\left(x_{2}\right)\cdots\mathcal{O}_{n}\left(x_{n}\right)\right\rangle =\left\langle \mathcal{O}_{1}\right\rangle \left\langle \mathcal{O}_{2}\right\rangle \cdots\left\langle \mathcal{O}_{n}\right\rangle,\label{eq:cluster}
\end{equation}
where on the right hand side we no longer need to specify the positions.

Since $\bar{Q}_{\dot{\alpha}A}$-exact objects decouple in the vacuum
expectation values, we can define an equivalence relation between
two chiral operators $\mathcal{O}_{1}\left(x_{1}\right)$ and $\mathcal{O}_{2}\left(x_{2}\right)$
if there exist a gauge invariant operator $X$ such that
\begin{equation}
\mathcal{O}_{1}\left(x_{1}\right)=\mathcal{O}_{2}\left(x_{2}\right)+\left[\bar{Q}_{\dot{\alpha}A},X\right].
\end{equation}
The set of equivalence classes of chiral operators forms a commutative
ring, known as the chiral ring \cite{Cachazo:2002ry}. We refer to
the generators of the chiral ring as basis chiral operators, whose
number is the complex dimension of the Coulomb branch. Any element
of the chiral ring can be represented uniquely in terms of the
basis element. We call the set of all such expressions the chiral
ring relations. The chiral ring is completely specified by the set
of basis chiral operators and the chiral ring relations. 

At the classical level, the chiral ring relations follow from group
theoretical identities. In this note, we focus on the $\mathrm{SU}(N)$
gauge theory with arbitrary matter. Let $\Phi$ be the adjoint scalar
field in the vector multiplet. We can choose the basis chiral operators to be 
\begin{equation}
\mathcal{O}_{n}=\frac{1}{n}\mathrm{Tr}\Phi^{n},\quad n=2,\cdots,N,
\end{equation}
whose vacuum expectation values parametrize the Coulomb branch of
moduli space of vacua. Since $\Phi$ is a traceless $N\times N$ matrix,
we can solve the Coulomb moduli $\vec{a}=\left(a_{1},\cdots,a_{N}\right)$
in terms of $\mathcal{O}_{n},n=2,\cdots,N$. Any other
chiral operator can be expressed in terms of $\vec{a}$. For example,
\begin{equation}
\mathcal{O}_{n_{1},n_{2},\cdots,n_{\ell}}=\prod_{i=1}^{\ell}\left(\frac{1}{n_{i}}\mathrm{Tr}\Phi^{n_{i}}\right)=\prod_{i=1}^{\ell}\left(\frac{1}{n_{i}}\sum_{\alpha=1}^{N}a_{\alpha}^{n_{i}}\right).
\end{equation}
We can get the classical chiral ring relations by substituting
$\vec{a}$ by $\mathcal{O}_{n},n=2,\cdots,N$.

A more elegant derivation of the classical chiral ring relation can
be given by expanding the characteristic polynomial of $\Phi$, 
\begin{eqnarray}
\det\left(x-\Phi\right) & = & x^{N}\exp\mathrm{Tr}\log\left(1-\frac{\Phi}{x}\right)\nonumber \\
 & = & x^{N}\exp\left(-\sum_{n=1}^{\infty}\frac{1}{nx^{n}}\mathrm{Tr}\Phi^{n}\right)\nonumber \\
 & = & x^{N}-x^{N-2}\mathcal{O}_{2}-x^{N-3}\mathcal{O}_{3}-x^{N-4}\left(\mathcal{O}_{4}-\frac{1}{2}\mathcal{O}_{2,2}\right)+\cdots.
\end{eqnarray}
Since the left-hand side is classically equal to the polynomial $\prod_{\alpha=1}^{N}\left(x-a_{\alpha}\right)$,
all the coefficients of negative powers of $x$ in the series expansion
must vanish. In this way, we directly obtain the classical chiral
ring relations. Notice that the classical chiral ring relations are
independent of details of the hypermultiplets. Quantum mechanically,
we can take the same set of basis chiral operators, but instanton
effects lead to important modifications of the chiral ring relations. 

\subsection{Chiral trace relations in the $\Omega$-background}

Since the $\Omega$-background breaks the translational symmetry of
$\mathbb{R}^{4}$, we can no longer use the position-independence
and the cluster decomposition to reduce the vacuum expectation value of a multi-trace chiral operator to those of single-trace
chiral operators. Consequently, we cannot avoid multi-trace chiral operators
in our discussion. 

A brute force way to derive the chiral trace relations in the $\Omega$-background
is as follows. One first computes the vacuum expectation values of
chiral operators $\mathcal{O}_{n}$ up to certain high order in (\ref{eq:vev}). Then
the Coulomb branch parameters $\vec{a}=\left(a_{1},\cdots,a_{N}\right)$ can
be solved in terms of $\mathcal{O}_{n},n=2,\cdots,N$,
with coefficients involving the masses $\vec{m}$, the instanton counting
parameter $\mathtt{q}$ and the $\Omega$-deformation parameters $\varepsilon_{1},\varepsilon_{2}$.
Substituting back to the expressions for other chiral operators, we
may then guess the chiral trace relations in the $\Omega$-background.
In practice, the computation can be extremely tedious, and we would like to 
look for a better approach.

Inspired by the derivation of the classical chiral ring relations
using the characteristic polynomial of $\Phi$, we consider the $\mathscr{Y}$-observable
\cite{Nekrasov:2015wsu},
\begin{equation}
\mathscr{Y}(x)=x^{N}\exp\left(-\sum_{n=1}^{\infty}\frac{1}{nx^{n}}\mathrm{Tr}\Phi^{n}\right),
\end{equation}
where we implicitly put $\Phi$ at the origin $0\in\mathbb{R}^{4}$,
which is the fixed point of the $\mathrm{SO}(4)$ rotation symmetry
of $\mathbb{R}^{4}$. At the classical level,
\begin{equation}
\mathscr{Y}(x)^{\mathrm{cl}}=x^{N}\exp\left(-\sum_{n=1}^{\infty}\frac{1}{nx^{n}}\sum_{\alpha=1}^{N}a_{\alpha}^{n}\right)=\prod_{\alpha=1}^{N}\left(x-a_{\alpha}\right).
\end{equation}
Hence $\mathscr{Y}(x)$ reduces to the characteristic polynomial $\det\left(x-\Phi\right)$
classically. 

In order to evaluate $\mathscr{Y}(x)$ at the instanton configuration $\vec{\lambda}$, we use the fact that \cite{Losev:2003py}
\begin{equation}
\left[\mathrm{Tr}e^{\beta\Phi}\right]_{\vec{\lambda}}
=\mathscr{E}_{\vec{\lambda}}
=\sum_{\alpha=1}^{N}\left[e^{\beta a_{\alpha}}-P\sum_{\square\in\lambda^{(\alpha)}}e^{\beta\left(a_{\alpha}+c_{\square}\right)}\right].\label{eq:ezPhi}
\end{equation}
Here $\mathscr{E}_{\vec{\lambda}}$ is the equivariant Chern character
of the universal bundle with the universal instanton connection over $\mathbb{C}^{2}\times\mathcal{M}_N$ evaluated at the fixed point $0\times\vec{\lambda}$, 
where $\mathcal{M}_N$ is the moduli space of framed $\mathrm{U}(N)$ instantons on $\mathbb{C}^{2}$. We define
\begin{equation}
P=\left(1-e^{\beta\varepsilon_{1}}\right)\left(1-e^{\beta\varepsilon_{2}}\right),
\end{equation}
and
\begin{equation}
c_{\square=\left(i,j\right)}=\varepsilon_{1}(i-1)+\varepsilon_{2}(j-1).
\end{equation}
We can compute $\left[\mathrm{Tr}\Phi^{n}\right]_{\vec{\lambda}}$
by expanding (\ref{eq:ezPhi}) in powers of $\beta$, and then we
have
\begin{equation}
\left[\mathscr{Y}(x)\right]_{\vec{\lambda}}=\prod_{\alpha=1}^{N}\left[\left(x-a_{\alpha}\right)\prod_{\square\in\lambda^{(\alpha)}}\frac{\left(x-a_{\alpha}-c_{\square}-\varepsilon_{1}\right)\left(x-a_{\alpha}-c_{\square}-\varepsilon_{2}\right)}{\left(x-a_{\alpha}-c_{\square}\right)\left(x-a_{\alpha}-c_{\square}-\varepsilon\right)}\right],
\end{equation}
where $\varepsilon=\varepsilon_{1}+\varepsilon_{2}$. It is convenient to write $\left[\mathscr{Y}(x)\right]_{\vec{\lambda}}$ succinctly as \cite{Nekrasov:2015wsu}
\begin{equation}
\left[\mathscr{Y}(x)\right]_{\vec{\lambda}}=\epsilon\left[e^{\beta x}\mathscr{E}_{\vec{\lambda}}^{\vee}\right] = (-1)^N \epsilon\left[e^{-\beta x}\mathscr{E}_{\vec{\lambda}}\right] , \label{Yepsilon}
\end{equation}
where $\epsilon$ is the conversion operator which maps characters into weights,
\begin{equation}
\epsilon\left\{ \sum_{i}n_{i}e^{\beta w_{i}}\right\} =\prod_{i}w_{i}^{n_{i}}, \label{epsilon}
\end{equation}
and $\vee$ is the dual operator,
\begin{equation}
\left(\sum_{i}n_{i}e^{\beta w_{i}}\right)^{\vee}=\sum_{i}n_{i}e^{- \beta w_{i}}. \label{vee}
\end{equation}
After carrying out many pairwise cancelations, $\left[\mathscr{Y}(x)\right]_{\vec{\lambda}}$ becomes
\begin{equation}
\left[\mathscr{Y}(x)\right]_{\vec{\lambda}}=\prod_{\alpha=1}^{N}\frac{\prod_{\boxplus\in\partial_{+}\lambda^{(\alpha)}}\left(x-a_{\alpha}-c_{\boxplus}\right)}{\prod_{\boxminus\in\partial_{-}\lambda^{(\alpha)}}\left(x-a_{\alpha}-c_{\boxminus}-\varepsilon\right)},
\end{equation}
where $\partial_{\pm}\lambda^{(\alpha)}$ are boxes that can be added
to or removed from $\lambda^{(\alpha)}$ while keeping the resulting
configuration a well-defined Young diagram. 

Unlike the characteristic polynomial, the function $\left[\mathscr{Y}(x)\right]_{\vec{\lambda}}$ has singularities in $x$.
Fortunately, it was found in \cite{Nekrasov:2015wsu} that the all-instanton information in the $\Omega$-background
can be encoded in a system of non-perturbative Dyson-Schwinger equations
\begin{equation}
\left\langle \mathscr{X}\left(\mathscr{Y}(x+\cdots)\right)\right\rangle =\mathscr{T}(x),
\end{equation}
where composite operators $\mathscr{X}(x)$ are built out of $\mathscr{Y}$-observables, and are called qq-characters. $\mathscr{T}(x)$ is a polynomial in $x$. Hence, there is no singularities in $\left\langle\mathscr{X}(x)\right\rangle$ for finite $x$. Although the $\mathscr{Y}$-observable
only depends on the vector multiplet, the qq-characters contain the information of the hypermultiplet.

Physically, a qq-character in an $\mathcal{N}=2$ supersymmetric gauge theory can be interpreted as the observable obtained by integrating out the auxiliary gauge theory living on a space transverse to the physical spacetime. As shown in \cite{Nekrasov:2016qym}, the moduli space of framed instantons on the combination of the physical spacetime and the auxiliary transverse space is compact when we turn on the appropriate $\Omega$-deformation and background $B$-field. Consequently, $\left\langle\mathscr{X}(x)\right\rangle$ cannot have singularities in $x$ since there is no phase transitions or no runaway flat directions at any special value of $x$.

In this paper, we mainly focus on the fundamental qq-character which takes the form
\begin{equation}
\mathscr{X}(x)=\mathscr{Y}(x)+\cdots.
\end{equation}
Here terms in $\cdots$ will cancel the poles of $\mathscr{Y}(x)$ in
the vacuum expectation value. In the weak coupling limit $\mathtt{q}\to0$, the fundamental qq-character 
$\mathscr{X}(x)\to\mathscr{Y}(x)$. We can follow the same logic
of the derivation of the classical chiral ring relations on $\mathbb{R}^{4}$
to directly obtain the exact chiral trace relations in the $\Omega$-background
by expanding $\mathscr{X}(x)$ around $x=\infty$ and requiring the
vacuum expectation values of $x^{-n}$ coefficients $\mathscr{X}^{(-n)}$
to be zero for all $n\in\mathbb{Z}^{+}$,
\begin{equation}
\left\langle \mathscr{X}^{(-n)}\right\rangle =0.
\end{equation}
 
\section{Basic examples of $\mathrm{SU}(N)$ gauge theories \label{sec:SU(N)}}

In this section, we illustrate our general strategy using two examples,
the pure $\mathrm{SU}(N)$ gauge theory, and the $\mathrm{SU}(N)$
gauge theory with $2N$ fundamental hypermultiplets. 

\subsection{Pure $\mathrm{SU}(N)$ theory}

\subsubsection{Partition function and qq-characters}

The partition function of the pure $\mathrm{SU}(N)$ gauge theory
in the $\Omega$-background depends on the instanton counting parameter
$\mathtt{q}=\Lambda^{2N}$, the Coulomb branch parameter $\vec{a}=\left(a_{1},\cdots,a_{N}\right)$
and the $\Omega$-deformation parameters $\varepsilon_{1},\varepsilon_{2}$. Applying the supersymmetric localization techniques, the partition function is given by the sum over a collection of
$N$ Young diagrams $\vec{\lambda}=\left(\lambda^{(1)},\cdots,\lambda^{(N)}\right)$
\cite{Nekrasov:2002qd,Nekrasov:2013xda,Nekrasov:2015wsu}, 
\begin{equation}
\mathcal{Z}\left(\vec{a},\mathtt{q};\varepsilon_{1},\varepsilon_{2}\right)
= \mathtt{q}^{-\frac{1}{2\varepsilon_{1}\varepsilon_{2}}\sum_{\alpha=1}^{N}a_{\alpha}^{2}}\sum_{\vec{\lambda}}\mathtt{q}^{\left|\vec{\lambda}\right|}\epsilon\left[\frac{\mathscr{E}_{\vec{\lambda}}\mathscr{E}_{\vec{\lambda}}^{\vee}}{P^{\vee}}\right],
\end{equation}
where $\left|\vec{\lambda}\right|$ is the total number of boxes in the $N$ Young diagrams $\vec{\lambda}$. Therefore, the measure factor is
\begin{equation}
\mu_{\vec{\lambda}}
= \mathtt{q}^{\left|\vec{\lambda}\right|-\frac{1}{2\varepsilon_{1}\varepsilon_{2}}\sum_{\alpha=1}^{N}a_{\alpha}^{2}}\epsilon\left[\frac{\mathscr{E}_{\vec{\lambda}}\mathscr{E}_{\vec{\lambda}}^{\vee}}{P^{\vee}}\right],
\end{equation}

In this simple example, it is easy to find the fundamental qq-character by comparing the measure factors $\mu_{\vec{\lambda}}$ and $\mu_{\vec{\lambda}^{\prime}}$, where $\vec{\lambda}^{\prime}$ is obtained from $\vec{\lambda}$
by removing a box $\blacksquare\in\partial_{-}\lambda^{(\alpha)}$ for
certain $\alpha\in \{1,\cdots,N\}$. Using the succinct notations (\ref{epsilon}) and (\ref{vee}), we have
\begin{equation}
\frac{\mu_{\vec{\lambda}^{\prime}}}{\mu_{\vec{\lambda}}}  = \mathtt{q}^{\left|\vec{\lambda}^{\prime}\right|-\left|\vec{\lambda}\right|}\epsilon\left[\frac{\mathscr{E}_{\vec{\lambda}^{\prime}}\mathscr{E}_{\vec{\lambda}^{\prime}}^{\vee}-\mathscr{E}_{\vec{\lambda}}\mathscr{E}_{\vec{\lambda}}^{\vee}}{P^{\vee}}\right].
\end{equation}
Since $\vec{\lambda}$ contains one more box than $\vec{\lambda}^{\prime}$,  we know that
\begin{equation}
\left|\vec{\lambda}^{\prime}\right|-\left|\vec{\lambda}\right|=-1.
\end{equation}
Meanwhile, according to (\ref{eq:ezPhi}), we have 
\begin{equation}
\mathscr{E}_{\vec{\lambda}^{\prime}} = \mathscr{E}_{\vec{\lambda}} + P \xi, \quad \xi= e^{\beta\left(a_{\alpha}+c_{\blacksquare}\right)},
\end{equation}
which gives
\begin{equation}
\mathscr{E}_{\vec{\lambda}^{\prime}}\mathscr{E}_{\vec{\lambda}^{\prime}}^{\vee}-\mathscr{E}_{\vec{\lambda}}\mathscr{E}_{\vec{\lambda}}^{\vee}
=  \mathscr{E}_{\vec{\lambda}^{\prime}} P^{\vee} \xi^{\vee} + \mathscr{E}_{\vec{\lambda}}^{\vee} P \xi.
\end{equation}
Therefore, we have
\begin{equation}
\frac{\mu_{\vec{\lambda}^{\prime}}}{\mu_{\vec{\lambda}}} = \mathtt{q}^{-1} \epsilon \left[\mathscr{E}_{\vec{\lambda}^{\prime}}\xi^{\vee}+e^{\beta\varepsilon}\mathscr{E}_{\vec{\lambda}}^{\vee}\xi \right].
\end{equation}
Now using (\ref{Yepsilon}), and the identity
\begin{eqnarray}
& & \mathrm{Res}_{x=a_{\alpha}+c_{\blacksquare}}\left[\mathscr{Y}\left(x+\varepsilon\right)\right]_{\vec{\lambda}} \nonumber\\
&=& \mathrm{Res}_{x=a_{\alpha}+c_{\blacksquare}} \left(\left[\mathscr{Y}\left(x+\varepsilon\right)\right]_{\vec{\lambda}^{\prime}} \frac{\left(x-a_{\alpha}-c_{\blacksquare}+\varepsilon_{1}\right)\left(x-a_{\alpha}-c_{\blacksquare}+\varepsilon_{2}\right)}{\left(x-a_{\alpha}-c_{\blacksquare}\right)\left(x-a_{\alpha}-c_{\blacksquare}+\varepsilon\right)} \right) \nonumber\\
&=& \frac{\varepsilon_{1}\varepsilon_{2}}{\varepsilon} \left[\mathscr{Y}\left(a_{\alpha}+c_{\blacksquare}+\varepsilon\right)\right]_{\vec{\lambda}^{\prime}},
\end{eqnarray}
we get
\begin{equation}
\mathrm{Res}_{x=a_{\alpha}+c_{\blacksquare}}\left(\left[\mathscr{Y}\left(x+\varepsilon\right)\right]_{\vec{\lambda}} \mu_{\vec{\lambda}} \right)
= \mathrm{Res}_{x=a_{\alpha}+c_{\blacksquare}}\left( (-1)^{N-1} \mathtt{q} \left[ \mathscr{Y}\left(x\right)^{-1}\right]_{\vec{\lambda}^{\prime}} \mu_{\vec{\lambda}^{\prime}} \right).
\end{equation}
As a result, we can take the fundamental qq-character of the theory to be
\begin{equation}
\mathscr{X}(x)=\mathscr{Y}(x+\varepsilon)+(-1)^{N}\mathtt{q}\mathscr{Y}(x)^{-1},
\end{equation}
whose vacuum expectation value $\left\langle \mathscr{X}(x)\right\rangle$ has no singularities in $x$, since the residues at all potential poles vanish due to pairwise cancellations. From the large $x$ behavior of $\mathscr{X}(x)$, we know that $\left\langle \mathscr{X}(x)\right\rangle$ is a polynomial in $x$.

\subsubsection{Chiral trace relations}

In the following, we shall write down explicitly the chiral trace
relations in the $\Omega$-background resulting from $\left\langle \mathscr{X}^{(-1)}\right\rangle =\left\langle \mathscr{X}^{(-2)}\right\rangle =\left\langle \mathscr{X}^{(-3)}\right\rangle =0$
for the gauge group $\mathrm{SU}(2)$ and $\mathrm{SU}(3)$. The generalizations
to higher orders or higher gauge groups are straightforward.

We denote
\begin{equation}
u_{n_{1},n_{2},\cdots,n_{\ell}}=\left\langle \mathcal{O}_{n_{1},n_{2},\cdots,n_{\ell}}\right\rangle .
\end{equation}
It is useful to notice from the structure of the partition function
that
\begin{equation}
\left\langle \mathcal{O}_{2}^{n}\mathcal{O}\right\rangle =\frac{1}{\mathcal{Z}}\left(-\varepsilon_{1}\varepsilon_{2}\mathtt{q}\frac{d}{d\mathtt{q}}\right)^{n}\Big(\left\langle \mathcal{O}\right\rangle \mathcal{Z}\Big)=\left(u_{2}-\varepsilon_{1}\varepsilon_{2}\mathtt{q}\frac{d}{d\mathtt{q}}\right)^{n}\left\langle \mathcal{O}\right\rangle .
\end{equation}
In particular, we have
\begin{eqnarray}
u_{2,2} & = & u_{2}^{2}-\varepsilon_{1}\varepsilon_{2}u_{2}^{\prime},\nonumber \\
u_{2,2,2} & = & u_{2}^{3}-3\varepsilon_{1}\varepsilon_{2}u_{2}^{\prime}u_{2}+\varepsilon_{1}^{2}\varepsilon_{2}^{2}u_{2}^{\prime\prime},
\end{eqnarray}
where we denote
\begin{equation}
u_{n}^{\prime}=\mathtt{q}\frac{d}{d\mathtt{q}}u_{n},\quad u_{n}^{\prime\prime}=\left(\mathtt{q}\frac{d}{d\mathtt{q}}\right)^{2}u_{n}.
\end{equation}

For the $\mathrm{SU}(2)$ theory, we have 
\begin{eqnarray}
\mathscr{X}^{(-1)} & = & -\mathcal{O}_{3},\nonumber \\
\mathscr{X}^{(-2)} & = & -\mathcal{O}_{4}+\frac{1}{2}\mathcal{O}_{2,2}+\mathtt{q}+\varepsilon\mathcal{O}_{3},\nonumber \\
\mathscr{X}^{(-3)} & = & -\mathcal{O}_{5}+\mathcal{O}_{2,3}+2\varepsilon\left(\mathcal{O}_{4}-\frac{1}{2}\mathcal{O}_{2,2}\right)-\varepsilon^{2}\mathcal{O}_{3}.
\end{eqnarray}
The non-perturbative Dyson-Schwinger equations $\left\langle \mathscr{X}^{(-1)}\right\rangle =\left\langle \mathscr{X}^{(-2)}\right\rangle =\left\langle \mathscr{X}^{(-3)}\right\rangle =0$
lead to
\begin{eqnarray}
u_{3} & = & 0,\nonumber \\
u_{4} & = & \frac{1}{2}\left(u_{2}^{2}-\varepsilon_{1}\varepsilon_{2}u_{2}^{\prime}\right)+\mathtt{q},\nonumber \\
u_{5} & = & 2\varepsilon\mathtt{q}.
\end{eqnarray}
We see that the classical chiral ring relations are modified both
by the instanton corrections (charactered by $\mathtt{q}$) and by
the $\Omega$-deformation (charactered by $\varepsilon_{1},\varepsilon_{2}$).
Our results provide an easy proof of the prediction made in \cite{Beccaria:2017rfz,Fachechi:2018bhe}
where they used the explicit localization results. The unavoidable
derivatives, which may seem peculiar in the original approach, appear
naturally in our approach.

Repeating the same kind of computation in the case of $\mathrm{SU}(3)$,
we find
\begin{eqnarray}
\mathscr{X}^{(-1)} & = & -\mathcal{O}_{4}+\frac{1}{2}\mathcal{O}_{2,2},\nonumber \\
\mathscr{X}^{(-2)} & = & -\mathcal{O}_{5}+\mathcal{O}_{2,3}+\varepsilon\left(\mathcal{O}_{4}-\frac{1}{2}\mathcal{O}_{2,2}\right),\nonumber \\
\mathscr{X}^{(-3)} & = & -\mathcal{O}_{6}+\mathcal{O}_{2,4}+\frac{1}{2}\mathcal{O}_{3,3}-\frac{1}{6}\mathcal{O}_{2,2,2}-\mathtt{q}\nonumber \\
 &  & +2\varepsilon\left(\mathcal{O}_{5}-\mathcal{O}_{2,3}\right)-\varepsilon^{2}\left(\mathcal{O}_{4}-\frac{1}{2}\mathcal{O}_{2,2}\right).
\end{eqnarray}
The non-perturbative Dyson-Schwinger equations $\left\langle \mathscr{X}^{(-1)}\right\rangle =\left\langle \mathscr{X}^{(-2)}\right\rangle =\left\langle \mathscr{X}^{(-3)}\right\rangle =0$
lead to
\begin{eqnarray}
u_{4} & = & \frac{1}{2}\left(u_{2}^{2}-\varepsilon_{1}\varepsilon_{2}u_{2}^{\prime}\right),\nonumber \\
u_{5} & = & u_{2}u_{3}-\varepsilon_{1}\varepsilon_{2}u_{3}^{\prime},\nonumber \\
u_{6} & = & \frac{1}{3}\left(u_{2}^{3}-3\varepsilon_{1}\varepsilon_{2}u_{2}^{\prime}u_{2}+\varepsilon_{1}^{2}\varepsilon_{2}^{2}u_{2}^{\prime\prime}\right)u_{2}^{3}+\frac{1}{2}u_{3,3}-\mathtt{q}.
\end{eqnarray}
Notice that in the $\Omega$-background, $u_{3,3}$ cannot be expressed
in terms of $u_{2}$ and $u_{3}$. Our result matches the prediction
made in \cite{Fachechi:2018bhe} perfectly. 

\subsection{$\mathrm{SU}(N)$ theory with $2N$ fundamental hypermultiplets \label{sec:fund}}

\subsubsection{Partition function and qq-characters}

The partition function of the $\mathrm{SU}(N)$ gauge theory with
$2N$ fundamental hypermultiplets in the $\Omega$-background is given
by \cite{Nekrasov:2002qd,Nekrasov:2013xda,Nekrasov:2015wsu}, 
\begin{equation}
\mathcal{Z}\left(\vec{a},\vec{m},\mathtt{q};\varepsilon_{1},\varepsilon_{2}\right) = 
\mathtt{q}^{-\frac{1}{2\varepsilon_{1}\varepsilon_{2}}\sum_{\alpha=1}^{N}a_{\alpha}^{2}}\sum_{\vec{\lambda}}\mathtt{q}^{\left|\vec{\lambda}\right|}
\epsilon\Bigg[\frac{\mathscr{E}_{\vec{\lambda}}\mathscr{E}_{\vec{\lambda}}^{\vee}-\mathscr{M}\mathscr{E}_{\vec{\lambda}}}{P^{\vee}}\Bigg],
\end{equation}
which depends on the Coulomb branch parameter $\vec{a}=\left(a_{1},\cdots,a_{N}\right)$,
the masses $\vec{m}=\left(m_{1},\cdots,m_{2N}\right)$, the instanton
counting parameter $\mathtt{q}=e^{2\pi\mathtt{i}\tau}$ with $\tau$
being the microscopic complexified coupling constant, and the $\Omega$-deformation
parameters $\varepsilon_{1},\varepsilon_{2}$. Here $\mathscr{M}$
encodes the information of masses
\begin{equation}
\mathscr{M}=\sum_{f=1}^{2N}e^{\beta m_{f}}.
\end{equation}

The qq-character of the theory can be obtained following the same
procedure as in the case of pure $\mathrm{SU}(N)$ gauge theory. In particular, we have
\begin{equation}
\frac{\mu_{\vec{\lambda}^{\prime}}}{\mu_{\vec{\lambda}}} = \mathtt{q}^{-1} \epsilon \left[\mathscr{E}_{\vec{\lambda}^{\prime}}\xi^{\vee}+e^{\beta\varepsilon}\mathscr{E}_{\vec{\lambda}}^{\vee}\xi - e^{\beta\varepsilon} \mathscr{M} \xi \right].
\end{equation}
Accordingly, the fundamental qq-character can be taken to be 
\begin{equation}
\mathscr{X}(x)=\mathscr{Y}(x+\varepsilon)+(-1)^{N}\mathtt{q}\mathscr{P}(x)\mathscr{Y}(x)^{-1},
\end{equation}
where
\begin{equation}
\mathscr{P}(x)=\prod_{f=1}^{2N}\left(x+m_{f}+\varepsilon\right)=x^{2N}+\sum_{n=1}^{2N}x^{2N-n}M_{n}.
\end{equation}
Again, $\left\langle \mathscr{X}(x)\right\rangle $ is a polynomial in $x$.

\subsubsection{Chiral trace relations}

Let us consider the gauge group $\mathrm{SU}(2)$. The first few terms
of $\mathscr{X}^{(-n)}$ are 
\begin{eqnarray}
\mathscr{X}^{(-1)} & = & -\left(1-\mathtt{q}\right)\mathcal{O}_{3}+\mathtt{q}M_{1}\mathcal{O}_{2}+\mathtt{q}M_{3},\nonumber \\
\mathscr{X}^{(-2)} & = & -\left(1-\mathtt{q}\right)\mathcal{O}_{4}+\left(\mathtt{q}M_{1}+\varepsilon\right)\mathcal{O}_{3}+\frac{1}{2}\left(1+\mathtt{q}\right)\mathcal{O}_{2,2}\nonumber \\
 &  & +\mathtt{q}M_{2}\mathcal{O}_{2}+\mathtt{q}M_{4},\nonumber \\
\mathscr{X}^{(-3)} & = & -\left(1-\mathtt{q}\right)\mathcal{O}_{5}+\left(\mathtt{q}M_{1}+2\varepsilon\right)\mathcal{O}_{4}+\left(1+\mathtt{q}\right)\mathcal{O}_{2,3}\nonumber \\
 &  & +\left(\mathtt{q}M_{2}-\varepsilon^{2}\right)\mathcal{O}_{3}+\left(\frac{1}{2}\mathtt{q}M_{1}-\varepsilon\right)\mathcal{O}_{2,2}+\mathtt{q}M_{3}\mathcal{O}_{2}.
\end{eqnarray}
Because of the non-perturbative Dyson-Schwinger equations, $\left\langle \mathscr{X}^{(-1)}\right\rangle =\left\langle \mathscr{X}^{(-2)}\right\rangle =\left\langle \mathscr{X}^{(-3)}\right\rangle =0$,
we have
\begin{eqnarray}
u_{3} & = & \frac{\mathtt{q}}{1-\mathtt{q}}\left(M_{1}u_{2}+M_{3}\right),\nonumber \\
u_{4} & = & \frac{\mathtt{q}M_{1}+\varepsilon}{1-\mathtt{q}}u_{3}+\frac{1+\mathtt{q}}{2\left(1-\mathtt{q}\right)}\left(u_{2}^{2}-\varepsilon_{1}\varepsilon_{2}u_{2}^{\prime}\right)+\frac{\mathtt{q}}{1-\mathtt{q}}M_{2}u_{2}+\frac{\mathtt{q}}{1-\mathtt{q}}M_{4}\nonumber \\
 & = & \frac{1+\mathtt{q}}{2\left(1-\mathtt{q}\right)}\left(u_{2}^{2}-\varepsilon_{1}\varepsilon_{2}u_{2}^{\prime}\right)+\left(\frac{\mathtt{q}\left(\mathtt{q}M_{1}+\varepsilon\right)M_{1}}{\left(1-\mathtt{q}\right)^{2}}+\frac{\mathtt{q}M_{2}}{1-\mathtt{q}}\right)u_{2}\nonumber \\
 &  & +\frac{\mathtt{q}\left(\mathtt{q}M_{1}+\varepsilon\right)M_{3}}{\left(1-\mathtt{q}\right)^{2}}+\frac{\mathtt{q}}{1-\mathtt{q}}M_{4},\nonumber \\
u_{5} & = & \frac{\mathtt{q}M_{1}+2\varepsilon}{1-\mathtt{q}}u_{4}+\frac{1+\mathtt{q}}{1-\mathtt{q}}\left(u_{2}u_{3}-\varepsilon_{1}\varepsilon_{2}u_{3}^{\prime}\right)\nonumber \\
 &  & +\frac{\mathtt{q}M_{2}-\varepsilon^{2}}{1-\mathtt{q}}u_{3}+\frac{\mathtt{q}M_{1}-2\varepsilon}{2\left(1-\mathtt{q}\right)}\left(u_{2}^{2}-\varepsilon_{1}\varepsilon_{2}u_{2}^{\prime}\right)+\frac{\mathtt{q}M_{3}}{1-\mathtt{q}}u_{2}\nonumber \\
 & = & \frac{\mathtt{q}\left(M_{1}+2\varepsilon\right)}{\left(1-\mathtt{q}\right)^{2}}\left(u_{2}^{2}-\varepsilon_{1}\varepsilon_{2}u_{2}^{\prime}\right)\nonumber \\
 &  & +\left(\frac{\mathtt{q}\left(\mathtt{q}M_{1}+\varepsilon\right)\left(\mathtt{q}M_{1}+2\varepsilon\right)M_{1}}{\left(1-\mathtt{q}\right)^{3}}+\frac{\mathtt{q}\left(\mathtt{q}M_{1}+2\varepsilon\right)M_{2}}{\left(1-\mathtt{q}\right)^{2}}+\frac{\mathtt{q}M_{3}}{1-\mathtt{q}}\right)u_{2}\nonumber \\
 &  & +\frac{\mathtt{q}^{2}\left(\mathtt{q}M_{1}+\varepsilon\right)\left(\mathtt{q}M_{1}+2\varepsilon\right)M_{3}}{\left(1-\mathtt{q}\right)^{3}}+\frac{\mathtt{q}\left(\mathtt{q}M_{1}+2\varepsilon\right)}{\left(1-\mathtt{q}\right)^{2}}M_{4}\nonumber \\
 &  & +\frac{1+\mathtt{q}}{1-\mathtt{q}}\left(u_{2}u_{3}-\varepsilon_{1}\varepsilon_{2}u_{3}^{\prime}\right)+\frac{\mathtt{q}M_{2}-\varepsilon^{2}}{1-\mathtt{q}}u_{3}.
\end{eqnarray}
Our result confirms the results obtained in \cite{Fucito:2015ofa}.
We see that the presence of the hypermultiplet significantly modifies
the chiral trace relations in the $\Omega$-background.

\section{The linear quiver gauge theory \label{sec:quiver}}

Let us briefly sketch how our method can be generalized to linear
quiver gauge theories.

The gauge group of the theory is
\begin{equation}
G=\prod_{i=1}^{r}\mathrm{SU}(N)_{i}.
\end{equation}
For each $\mathrm{SU}(N)_{i}$ factor, we have a vector multiplet
transforming in the adjoint representation of $\mathrm{SU}(N)_{i}$.
The matter content of the theory spits into $N$ hypermultiplets in
the anti-fundamental representation of $\mathrm{SU}(N)_{1}$, one
hypermultiplet in the bifundamental representation of $\mathrm{SU}(N)_{i}\times\mathrm{SU}(N)_{i+1}$
for $i=1,\cdots,r-1$, and $N$ fundamental representation of $\mathrm{SU}(N)_{r}$.
The action of the theory is parametrized by the instanton counting
parameters
\begin{equation}
\underline{\mathtt{q}}=\left\{ \mathtt{q}_{i}=e^{2\pi\mathtt{i}\tau_{i}},i=1,\cdots r\right\} ,
\end{equation}
where $\tau_{i}$ is the complexified gauge coupling for $\mathrm{SU}(N)_{i}$.
It is convenient to extend the gauge group $\prod_{i=1}^{r}\mathrm{SU}(N)_{i}$
to $\prod_{i=0}^{r+1}\mathrm{U}(N)_{i}$ if we set $\mathtt{q}_{0}=\mathtt{q}_{r+1}=0$.
All the masses are formally viewed as Coulomb branch parameters, leading
to the extended moduli space with parameters
\begin{equation}
\left(a_{i,\alpha}\right)_{\alpha=1}^{N},\quad i=0,\cdots,r+1.
\end{equation}
The physical Coulomb branch parameters are given by
\begin{equation}
\mathtt{a}_{i,\alpha}=a_{i,\alpha}-\frac{1}{N}\sum_{\alpha=1}^{N}a_{i,\alpha},\quad i=1,\cdots,r.
\end{equation}
We introduce $z_{0},z_{1},\cdots,z_{r+1}$, such that $z_{-1}=\infty$,
$z_{r+1}=0$, and
\begin{equation}
\mathtt{q}_{i}=\frac{z_{i}}{z_{i-1}},\quad i=1,\cdots,r.
\end{equation}
We can define $\mathscr{Y}$-observables for $i=0$ and $i=r+1$ as
\begin{equation}
\mathscr{Y}_{0}(x)=\prod_{\alpha=1}^{N}\left(x-a_{0,\alpha}\right),\quad\mathscr{Y}_{r+1}(x)=\prod_{\alpha=1}^{N}\left(x-a_{r+1,\alpha}\right).
\end{equation}

The fundamental qq-characters are given by \cite{Nekrasov:2015wsu}
\begin{eqnarray}
\mathscr{X}_{\ell}(x) & = & \frac{\mathscr{Y}_{0}\left(x+\varepsilon\left(1-\ell\right)\right)}{z_{0}z_{1}\cdots z_{\ell-1}}\sum_{\substack{I\subset[0,r]\\
|I|=\ell
}
}\prod_{i\in I}\left[z_{i}\Xi_{i}\left(x+\varepsilon\left(h_{I}(i)+1-\ell\right)\right)\right]\nonumber \\
 & = & \mathscr{Y}_{\ell}\left(x+\varepsilon\right)+\cdots,\quad\ell=0,1,\cdots,r+1,
\end{eqnarray}
where $[0,r]=\left\{ 0,1,2,\cdots,r\right\} $, the function $h_{I}(i)$
is the number of elements in the set $I$ which is less than $i$,
and 
\begin{equation}
\Xi_{i}(x)=\frac{\mathscr{Y}{}_{i+1}(x+\varepsilon)}{\mathscr{Y}{}_{i}(x)},\quad i=0,\cdots,r.
\end{equation}
In order to deal with all of the fundamental qq-characters at the
same time, we introduce the generating function
\begin{eqnarray}
G_{r}(x;t) & = & \Delta_{r}^{-1}\mathscr{Y}_{0}(x)^{-1}\sum_{\ell=0}^{r+1}z_{0}z_{1}\cdots z_{\ell-1}t^{\ell}\mathscr{X}_{\ell}\left(x-\varepsilon(1-\ell)\right)\nonumber \\
 & = & \Delta_{r}^{-1}\sum_{I\subset[0,r]}\left[\left(\prod_{i\in I}tz_{i}\right)\prod_{i\in I}\Xi_{i}\left(x+\varepsilon h_{I}(i)\right)\right],\label{eq:Gr}
\end{eqnarray}
where
\begin{equation}
\Delta_{r}=\prod_{i=0}^{r}\left(1+tz_{i}\right)=\sum_{I\subset[0,r]}\left(\prod_{i\in I}tz_{i}\right)
\end{equation}
is a normalization factor. We can find a relation between $G_{r}(x;t)$
and $G_{r-1}(x;t)$ by classifying the set $I\subset[0,r]$ depending
on whether $r\in I$,
\begin{eqnarray}
G_{r}(x;t) & = & \Delta_{r}^{-1}\sum_{I^{\prime}\subset[0,r-1]}\left[\left(\prod_{i\in I^{\prime}}tz_{i}\right)\prod_{i\in I^{\prime}}\Xi_{i}\left(x+\varepsilon h_{I^{\prime}}(i)\right)\right]\nonumber \\
 &  & +\Delta_{r}^{-1}tz_{r}\sum_{I^{\prime}\subset[0,r-1]}\left[\left(\prod_{i\in I^{\prime}}tz_{i}\right)\Xi_{r}\left(x+\varepsilon|I^{\prime}|\right)\prod_{i\in I^{\prime}}\Xi_{i}\left(x+\varepsilon h_{I^{\prime}}(i)\right)\right]\nonumber \\
 & = & \frac{1}{1+tz_{r}}G_{r-1}(x;t)+\frac{tz_{r}}{1+tz_{r}}\Delta_{r-1}^{-1}\Xi_{r}\left(x+\varepsilon t\frac{\partial}{\partial t}\right)\left(\Delta_{r-1}G_{r-1}(x;t)\right).\label{eq:Grec}
\end{eqnarray}
The large $x$ expansion of the function $\Xi_{i}(x)$ is 
\begin{equation}
\Xi_{i}(x)=\exp\sum_{n=1}^{\infty}\left(\frac{\mathcal{O}_{n}^{(i)}}{x^{n}}-\frac{\mathcal{O}_{n}^{(i+1)}}{\left(x+\varepsilon\right)^{n}}\right)=\sum_{n=0}^{\infty}\frac{\zeta_{i,n}}{x^{n}},
\end{equation}
where 
\begin{equation}
\mathcal{O}_{n}^{(i)}=\frac{1}{n}\mathrm{Tr}\Phi_{i}^{n},\quad i=0,\cdots,r+1.
\end{equation}
The first few terms of $\zeta_{i,n}$ are
\begin{eqnarray}
\zeta_{i,0} & = & 1\nonumber \\
\zeta_{i,1} & = & \mathcal{O}_{1}^{(i)}-\mathcal{O}_{1}^{(i+1)},\nonumber \\
\zeta_{i,2} & = & \mathcal{O}_{2}^{(i)}-\mathcal{O}_{2}^{(i+1)}+\varepsilon\mathcal{O}_{1}^{(i+1)}+\frac{1}{2}\left(\mathcal{O}_{1}^{(i)}-\mathcal{O}_{1}^{(i+1)}\right)^{2},\nonumber \\
\zeta_{i,3} & = & \mathcal{O}_{3}^{(i)}-\mathcal{O}_{3}^{(i+1)}+\left(\mathcal{O}_{1}^{(i)}-\mathcal{O}_{1}^{(i+1)}\right)\left(\mathcal{O}_{2}^{(i)}-\mathcal{O}_{2}^{(i+1)}\right)\nonumber \\
 &  & +\left(\mathcal{O}_{1}^{(i)}-\mathcal{O}_{1}^{(i+1)}\right)^{3}+\varepsilon\left(\mathcal{O}_{1}^{(i)}-\mathcal{O}_{1}^{(i+1)}\right)\mathcal{O}_{1}^{(i+1)}\nonumber \\
 &  & +2\varepsilon\mathcal{O}_{2}^{(i+1)}-\varepsilon^{2}\mathcal{O}_{1}^{(i+1)}.\label{eq:zeta}
\end{eqnarray}
We can then expand $G_{r}(x;t)$ as 
\begin{equation}
G_{r}(x;t)=\sum_{n=0}^{\infty}\frac{G_{r}^{(n)}(t)}{x^{n}},
\end{equation}
and use (\ref{eq:Grec}) to compute the expansion coefficients $G_{r}^{(n)}(t)$
recursively. The first few terms are given by \cite{Jeong:2017mfh}
\begin{eqnarray}
G_{r}^{(1)} & = & U_{r}[1],\nonumber \\
G_{r}^{(2)} & = & U_{r}[2]+U_{r}[1,1]-\varepsilon U_{r}[0,1],\nonumber \\
G_{r}^{(3)} & = & U_{r}[3]+U_{r}[2,1]+U_{r}[1,2]-\varepsilon\left(U_{r}[1,1]+2U_{r}[0,2]\right)+\varepsilon^{2}U_{r}[0,1]\nonumber \\
 &  & +U_{r}[1,1,1]-\varepsilon\left(2U_{r}[0,1,1]+U_{r}[1,0,1]\right)+2\varepsilon^{2}U_{r}[0,0,1],\label{eq:GU}
\end{eqnarray}
where $U_{r}\left[s_{1},s_{2},\cdots,s_{\ell}\right]$ is defined
for non-negative integers $\left[s_{1},\cdots,s_{\ell}\right]$ as
\begin{equation}
U_{r}\left[s_{1},s_{2},\cdots,s_{\ell}\right]=\sum_{0\leq i_{1}<\cdots<i_{\ell}\leq r}\prod_{n=1}^{\ell}\left(\frac{tz_{i_{n}}}{1+tz_{i_{n}}}\zeta_{i_{n},s_{n}}\right).\label{eq:Ur}
\end{equation}

The non-perturbative Dyson-Schwinger equations result in the relations
\begin{equation}
0=\left\langle \left(\mathscr{Y}_{0}(x)G_{r}(x;t)\right)^{(-n)}\right\rangle =\sum_{n=0}^{N}\mathcal{A}_{j}\left\langle G_{r}^{(N+n-j)}(t)\right\rangle ,\quad n\in\mathbb{Z}^{+},\label{eq:DSquiver}
\end{equation}
where
\begin{equation}
\mathscr{Y}_{0}(x)=\prod_{\alpha=1}^{N}\left(x-a_{0,\alpha}\right)=\sum_{j=0}^{N}\mathcal{A}_{j}x^{N-j}.
\end{equation}
After expressing $G_{r}^{(n)}(t)$ entirely in terms of chiral operators
using (\ref{eq:zeta})(\ref{eq:GU})(\ref{eq:Ur}), we finally derive
the chiral trace relations in the $\Omega$-background from (\ref{eq:DSquiver}). 
\acknowledgments
We are grateful to Nikita Nekrasov and Wenbin Yan for helpful discussions. SJ and XZ are supported by the US Department of Energy under grant DE-SC0010008. The work of SJ was also supported by the NSF grant PHY 1404446 and by the Simons Center for Geometry and Physics.

\bibliographystyle{JHEP}
\bibliography{Ref}

\providecommand{\href}[2]{#2}\begingroup\raggedright\begin{thebibliography}{10}

\bibitem{Seiberg:1994rs}
N.~Seiberg and E.~Witten, \emph{{Electric - magnetic duality, monopole
  condensation, and confinement in N=2 supersymmetric Yang-Mills theory}},
  \href{http://dx.doi.org/10.1016/0550-3213(94)90124-4,
  10.1016/0550-3213(94)00449-8}{\emph{Nucl. Phys.} {\bfseries B426} (1994)
  19--52}, [\href{https://arxiv.org/abs/hep-th/9407087}{{\ttfamily
  hep-th/9407087}}].

\bibitem{Seiberg:1994aj}
N.~Seiberg and E.~Witten, \emph{{Monopoles, duality and chiral symmetry
  breaking in N=2 supersymmetric QCD}},
  \href{http://dx.doi.org/10.1016/0550-3213(94)90214-3}{\emph{Nucl. Phys.}
  {\bfseries B431} (1994) 484--550},
  [\href{https://arxiv.org/abs/hep-th/9408099}{{\ttfamily hep-th/9408099}}].

\bibitem{Moore:1997dj}
G.~W. Moore, N.~Nekrasov and S.~Shatashvili, \emph{{Integrating over Higgs
  branches}}, \href{http://dx.doi.org/10.1007/PL00005525}{\emph{Commun. Math.
  Phys.} {\bfseries 209} (2000) 97--121},
  [\href{https://arxiv.org/abs/hep-th/9712241}{{\ttfamily hep-th/9712241}}].

\bibitem{Nekrasov:2002qd}
N.~A. Nekrasov, \emph{{Seiberg-Witten prepotential from instanton counting}},
  \href{http://dx.doi.org/10.4310/ATMP.2003.v7.n5.a4}{\emph{Adv. Theor. Math.
  Phys.} {\bfseries 7} (2003) 831--864},
  [\href{https://arxiv.org/abs/hep-th/0206161}{{\ttfamily hep-th/0206161}}].

\bibitem{Nekrasov:2003rj}
N.~Nekrasov and A.~Okounkov, \emph{{Seiberg-Witten theory and random
  partitions}}, \href{http://dx.doi.org/10.1007/0-8176-4467-9_15}{\emph{Prog.
  Math.} {\bfseries 244} (2006) 525--596},
  [\href{https://arxiv.org/abs/hep-th/0306238}{{\ttfamily hep-th/0306238}}].

\bibitem{Shadchin:2004yx}
S.~Shadchin, \emph{{Saddle point equations in Seiberg-Witten theory}},
  \href{http://dx.doi.org/10.1088/1126-6708/2004/10/033}{\emph{JHEP} {\bfseries
  10} (2004) 033}, [\href{https://arxiv.org/abs/hep-th/0408066}{{\ttfamily
  hep-th/0408066}}].

\bibitem{Shadchin:2005cc}
S.~Shadchin, \emph{{Cubic curves from instanton counting}},
  \href{http://dx.doi.org/10.1088/1126-6708/2006/03/046}{\emph{JHEP} {\bfseries
  03} (2006) 046}, [\href{https://arxiv.org/abs/hep-th/0511132}{{\ttfamily
  hep-th/0511132}}].

\bibitem{Nekrasov:2012xe}
N.~Nekrasov and V.~Pestun, \emph{{Seiberg-Witten geometry of four dimensional
  N=2 quiver gauge theories}},
  \href{https://arxiv.org/abs/1211.2240}{{\ttfamily 1211.2240}}.

\bibitem{Zhang:2019msw}
X.~Zhang, \emph{{Seiberg-Witten geometry of four-dimensional $N$=2 SO--USp
  quiver gauge theories}},
  \href{http://dx.doi.org/10.1103/PhysRevD.100.125015}{\emph{Phys. Rev.}
  {\bfseries D100} (2019) 125015},
  [\href{https://arxiv.org/abs/1910.10104}{{\ttfamily 1910.10104}}].

\bibitem{Losev:2003py}
A.~S. Losev, A.~Marshakov and N.~A. Nekrasov, \emph{{Small instantons, little
  strings and free fermions}},
  \href{https://arxiv.org/abs/hep-th/0302191}{{\ttfamily hep-th/0302191}}.

\bibitem{Marshakov:2006ii}
A.~Marshakov and N.~Nekrasov, \emph{{Extended Seiberg-Witten Theory and
  Integrable Hierarchy}},
  \href{http://dx.doi.org/10.1088/1126-6708/2007/01/104}{\emph{JHEP} {\bfseries
  01} (2007) 104}, [\href{https://arxiv.org/abs/hep-th/0612019}{{\ttfamily
  hep-th/0612019}}].

\bibitem{Nekrasov:2009zza}
N.~A. Nekrasov, \emph{{Two-dimensional topological strings revisited}},
  \href{http://dx.doi.org/10.1007/s11005-009-0312-9}{\emph{Lett. Math. Phys.}
  {\bfseries 88} (2009) 207--253}.

\bibitem{Zhang:2016xqi}
X.~Zhang, \emph{{Partition function of $\mathcal{N}=2$ supersymmetric gauge
  theory and two-dimensional Yang-Mills theory}},
  \href{http://dx.doi.org/10.1103/PhysRevD.96.025008}{\emph{Phys. Rev.}
  {\bfseries D96} (2017) 025008},
  [\href{https://arxiv.org/abs/1609.09050}{{\ttfamily 1609.09050}}].

\bibitem{Gaiotto:2009we}
D.~Gaiotto, \emph{{N=2 dualities}},
  \href{http://dx.doi.org/10.1007/JHEP08(2012)034}{\emph{JHEP} {\bfseries 08}
  (2012) 034}, [\href{https://arxiv.org/abs/0904.2715}{{\ttfamily 0904.2715}}].

\bibitem{Gaiotto:2009hg}
D.~Gaiotto, G.~W. Moore and A.~Neitzke, \emph{{Wall-crossing, Hitchin Systems,
  and the WKB Approximation}},
  \href{https://arxiv.org/abs/0907.3987}{{\ttfamily 0907.3987}}.

\bibitem{Alday:2009aq}
L.~F. Alday, D.~Gaiotto and Y.~Tachikawa, \emph{{Liouville Correlation
  Functions from Four-dimensional Gauge Theories}},
  \href{http://dx.doi.org/10.1007/s11005-010-0369-5}{\emph{Lett. Math. Phys.}
  {\bfseries 91} (2010) 167--197},
  [\href{https://arxiv.org/abs/0906.3219}{{\ttfamily 0906.3219}}].

\bibitem{Wyllard:2009hg}
N.~Wyllard, \emph{{A(N-1) conformal Toda field theory correlation functions
  from conformal N = 2 SU(N) quiver gauge theories}},
  \href{http://dx.doi.org/10.1088/1126-6708/2009/11/002}{\emph{JHEP} {\bfseries
  11} (2009) 002}, [\href{https://arxiv.org/abs/0907.2189}{{\ttfamily
  0907.2189}}].

\bibitem{Nekrasov:2009rc}
N.~A. Nekrasov and S.~L. Shatashvili, \emph{{Quantization of Integrable Systems
  and Four Dimensional Gauge Theories}},  in \emph{{Proceedings, 16th
  International Congress on Mathematical Physics (ICMP09): Prague, Czech
  Republic, August 3-8, 2009}}, pp.~265--289, 2009.
\newblock \href{https://arxiv.org/abs/0908.4052}{{\ttfamily 0908.4052}}.
\newblock \href{http://dx.doi.org/10.1142/9789814304634_0015}{DOI}.

\bibitem{Nekrasov:2011bc}
N.~Nekrasov, A.~Rosly and S.~Shatashvili, \emph{{Darboux coordinates, Yang-Yang
  functional, and gauge theory}},
  \href{http://dx.doi.org/10.1016/j.nuclphysbps.2011.04.150}{\emph{Nucl. Phys.
  Proc. Suppl.} {\bfseries 216} (2011) 69--93},
  [\href{https://arxiv.org/abs/1103.3919}{{\ttfamily 1103.3919}}].

\bibitem{Nekrasov:2013xda}
N.~Nekrasov, V.~Pestun and S.~Shatashvili, \emph{{Quantum geometry and quiver
  gauge theories}},
  \href{http://dx.doi.org/10.1007/s00220-017-3071-y}{\emph{Commun. Math. Phys.}
  {\bfseries 357} (2018) 519--567},
  [\href{https://arxiv.org/abs/1312.6689}{{\ttfamily 1312.6689}}].

\bibitem{Cachazo:2002ry}
F.~Cachazo, M.~R. Douglas, N.~Seiberg and E.~Witten, \emph{{Chiral rings and
  anomalies in supersymmetric gauge theory}},
  \href{http://dx.doi.org/10.1088/1126-6708/2002/12/071}{\emph{JHEP} {\bfseries
  12} (2002) 071}, [\href{https://arxiv.org/abs/hep-th/0211170}{{\ttfamily
  hep-th/0211170}}].

\bibitem{Cachazo:2003yc}
F.~Cachazo, N.~Seiberg and E.~Witten, \emph{{Chiral rings and phases of
  supersymmetric gauge theories}},
  \href{http://dx.doi.org/10.1088/1126-6708/2003/04/018}{\emph{JHEP} {\bfseries
  04} (2003) 018}, [\href{https://arxiv.org/abs/hep-th/0303207}{{\ttfamily
  hep-th/0303207}}].

\bibitem{Beccaria:2017rfz}
M.~Beccaria, A.~Fachechi and G.~Macorini, \emph{{Chiral trace relations in
  $\Omega$-deformed $ \mathcal{N}=2 $ theories}},
  \href{http://dx.doi.org/10.1007/JHEP05(2017)023}{\emph{JHEP} {\bfseries 05}
  (2017) 023}, [\href{https://arxiv.org/abs/1702.01254}{{\ttfamily
  1702.01254}}].

\bibitem{Fachechi:2018bhe}
A.~Fachechi, G.~Macorini and M.~Beccaria, \emph{{Chiral trace relations in
  $\Omega$-deformed $\mathcal N=2$ theories}},
  \href{http://dx.doi.org/10.1088/1742-6596/965/1/012013}{\emph{J. Phys. Conf.
  Ser.} {\bfseries 965} (2018) 012013}.

\bibitem{Alba:2010qc}
V.~A. Alba, V.~A. Fateev, A.~V. Litvinov and G.~M. Tarnopolskiy, \emph{{On
  combinatorial expansion of the conformal blocks arising from AGT
  conjecture}}, \href{http://dx.doi.org/10.1007/s11005-011-0503-z}{\emph{Lett.
  Math. Phys.} {\bfseries 98} (2011) 33--64},
  [\href{https://arxiv.org/abs/1012.1312}{{\ttfamily 1012.1312}}].

\bibitem{Fateev:2011hq}
V.~A. Fateev and A.~V. Litvinov, \emph{{Integrable structure, W-symmetry and
  AGT relation}}, \href{http://dx.doi.org/10.1007/JHEP01(2012)051}{\emph{JHEP}
  {\bfseries 01} (2012) 051},
  [\href{https://arxiv.org/abs/1109.4042}{{\ttfamily 1109.4042}}].

\bibitem{Fucito:2015ofa}
F.~Fucito, J.~F. Morales and R.~Poghossian, \emph{{Wilson loops and chiral
  correlators on squashed spheres}},
  \href{http://dx.doi.org/10.1007/JHEP11(2015)064}{\emph{JHEP} {\bfseries 11}
  (2015) 064}, [\href{https://arxiv.org/abs/1507.05426}{{\ttfamily
  1507.05426}}].

\bibitem{Ashok:2016ewb}
S.~K. Ashok, M.~Billo, E.~Dell'Aquila, M.~Frau, A.~Lerda, M.~Moskovic et~al.,
  \emph{{Chiral observables and S-duality in N = 2* U(N) gauge theories}},
  \href{http://dx.doi.org/10.1007/JHEP11(2016)020}{\emph{JHEP} {\bfseries 11}
  (2016) 020}, [\href{https://arxiv.org/abs/1607.08327}{{\ttfamily
  1607.08327}}].

\bibitem{Nekrasov:2015wsu}
N.~Nekrasov, \emph{{BPS/CFT correspondence: non-perturbative Dyson-Schwinger
  equations and qq-characters}},
  \href{http://dx.doi.org/10.1007/JHEP03(2016)181}{\emph{JHEP} {\bfseries 03}
  (2016) 181}, [\href{https://arxiv.org/abs/1512.05388}{{\ttfamily
  1512.05388}}].

\bibitem{Nekrasov:2017gzb}
N.~Nekrasov, \emph{{BPS/CFT correspondence V: BPZ and KZ equations from
  qq-characters}},  \href{https://arxiv.org/abs/1711.11582}{{\ttfamily
  1711.11582}}.

\bibitem{Jeong:2017mfh}
S.~Jeong and X.~Zhang, \emph{{BPZ equations for higher degenerate fields and
  non-perturbative Dyson-Schwinger equations}},
  \href{https://arxiv.org/abs/1710.06970}{{\ttfamily 1710.06970}}.

\bibitem{Nekrasov:2017rqy}
N.~Nekrasov, \emph{{BPS/CFT correspondence IV: sigma models and defects in
  gauge theory}},
  \href{http://dx.doi.org/10.1007/s11005-018-1115-7}{\emph{Lett. Math. Phys.}
  {\bfseries 109} (2019) 579--622},
  [\href{https://arxiv.org/abs/1711.11011}{{\ttfamily 1711.11011}}].

\bibitem{Jeong:2018qpc}
S.~Jeong and N.~Nekrasov, \emph{{Opers, surface defects, and Yang-Yang
  functional}},  \href{https://arxiv.org/abs/1806.08270}{{\ttfamily
  1806.08270}}.

\bibitem{Jeong:2017pai}
S.~Jeong, \emph{{Splitting of surface defect partition functions and integrable
  systems}},
  \href{http://dx.doi.org/10.1016/j.nuclphysb.2018.12.007}{\emph{Nucl. Phys.}
  {\bfseries B938} (2019) 775--806},
  [\href{https://arxiv.org/abs/1709.04926}{{\ttfamily 1709.04926}}].

\bibitem{Bourgine:2017jsi}
J.-E. Bourgine, M.~Fukuda, K.~Harada, Y.~Matsuo and R.-D. Zhu, \emph{{(p,
  q)-webs of DIM representations, 5d $ \mathcal{N}=1 $ instanton partition
  functions and qq-characters}},
  \href{http://dx.doi.org/10.1007/JHEP11(2017)034}{\emph{JHEP} {\bfseries 11}
  (2017) 034}, [\href{https://arxiv.org/abs/1703.10759}{{\ttfamily
  1703.10759}}].

\bibitem{Bourgine:2019phm}
J.-E. Bourgine and S.~Jeong, \emph{{New quantum toroidal algebras from 5D
  $\mathcal{N}=1$ instantons on orbifolds}},
  \href{https://arxiv.org/abs/1906.01625}{{\ttfamily 1906.01625}}.

\bibitem{Pestun:2007rz}
V.~Pestun, \emph{{Localization of gauge theory on a four-sphere and
  supersymmetric Wilson loops}},
  \href{http://dx.doi.org/10.1007/s00220-012-1485-0}{\emph{Commun. Math. Phys.}
  {\bfseries 313} (2012) 71--129},
  [\href{https://arxiv.org/abs/0712.2824}{{\ttfamily 0712.2824}}].

\bibitem{Hama:2012bg}
N.~Hama and K.~Hosomichi, \emph{{Seiberg-Witten Theories on Ellipsoids}},
  \href{http://dx.doi.org/10.1007/JHEP09(2012)033,
  10.1007/JHEP10(2012)051}{\emph{JHEP} {\bfseries 09} (2012) 033},
  [\href{https://arxiv.org/abs/1206.6359}{{\ttfamily 1206.6359}}].

\bibitem{Nekrasov:2016qym}
N.~Nekrasov, \emph{{BPS/CFT correspondence II: Instantons at crossroads, moduli
  and compactness theorem}},
  \href{http://dx.doi.org/10.4310/ATMP.2017.v21.n2.a4}{\emph{Adv. Theor. Math.
  Phys.} {\bfseries 21} (2017) 503--583},
  [\href{https://arxiv.org/abs/1608.07272}{{\ttfamily 1608.07272}}].

\end{thebibliography}\endgroup

\end{document}